\documentclass[aps,pra,twocolumn,10pt,showpacs,amsmath,amssymb,superscriptaddress]{revtex4}
\usepackage{mathrsfs}
\usepackage{graphicx}
\usepackage{amsmath}
\usepackage{amssymb}
\usepackage{amsfonts}
\usepackage{color}
\usepackage{CJK}

\def\hatd#1{\hat{#1}^\dagger}
\def\bra#1{\left\langle{#1}\right|}
\def\ket#1{\left|{#1}\right\rangle}
\def\braket#1#2{\left\langle{{#1}}\mathrel{\left|{\vphantom{{#1}{#2}}}\right.\kern-\nulldelimiterspace}{{#2}}\right\rangle}

\begin{document}

\title{Cluster mean-field signature of entanglement entropy in bosonic superfluid-insulator transitions}

\author{Li Zhang}
\affiliation{TianQin Research Center \& School of Physics and Astronomy, Sun Yat-Sen University (Zhuhai Campus), Zhuhai 519082, China}
\affiliation{State Key Laboratory of Optoelectronic Materials and Technologies, Sun Yat-Sen University (Guangzhou Campus), Guangzhou 510275, China}

\author{Xizhou Qin}
\affiliation{TianQin Research Center \& School of Physics and Astronomy, Sun Yat-Sen University (Zhuhai Campus), Zhuhai 519082, China}

\author{Yongguan Ke}
\affiliation{TianQin Research Center \& School of Physics and Astronomy, Sun Yat-Sen University (Zhuhai Campus), Zhuhai 519082, China}
\affiliation{State Key Laboratory of Optoelectronic Materials and Technologies, Sun Yat-Sen University (Guangzhou Campus), Guangzhou 510275, China}

\author{Chaohong Lee}
\altaffiliation{Email: lichaoh2@mail.sysu.edu.cn; chleecn@gmail.com}
\affiliation{TianQin Research Center \& School of Physics and Astronomy, Sun Yat-Sen University (Zhuhai Campus), Zhuhai 519082, China}
\affiliation{State Key Laboratory of Optoelectronic Materials and Technologies, Sun Yat-Sen University (Guangzhou Campus), Guangzhou 510275, China}
\affiliation{Synergetic Innovation Center for Quantum Effects and Applications, Hunan Normal University, Changsha 410081, China}

\date{\today}

\begin{abstract}
  Entanglement entropy (EE), a fundamental conception in quantum information for characterizing entanglement, has been extensively employed to explore quantum phase transitions (QPTs).
  Although the conventional single-site mean-field (MF) approach successfully predicts the emergence of QPTs, it fails to include any entanglement.
  Here, for the first time, in the framework of a cluster MF treatment, we extract the signature of EE in the bosonic superfluid-insulator (SI) transitions.
  We consider a trimerized Kagom\'e lattice of interacting bosons, in which each trimer is treated as a cluster, and implement the cluster MF treatment by decoupling all inter-trimer hopping.
  In addition to superfluid and integer insulator phases, we find that fractional insulator phases appear when the tunneling is dominated by the intra-trimer part.
  To quantify the residual bipartite entanglement in a cluster, we calculate the second-order R\'enyi entropy, which can be experimentally measured by quantum interference of many-body twins.
  The second-order R\'enyi entropy itself is continuous everywhere, however, the continuousness of its first-order derivative breaks down at the phase boundary.
  This means that the bosonic SI transitions can still be efficiently captured by the residual entanglement in our cluster MF treatment.
  Besides to the bosonic SI transitions, our cluster MF treatment may also be used to capture the signature of EE for other QPTs in quantum superlattice models.
\end{abstract}


\maketitle

\section{introduction\label{Sec1}}

Quantum entanglement~\cite{Horodecki2009,Nicolas2015,Carmeli2016} an essential resource enabling modern quantum technologies, has a broad impact in understanding quantum phase transitions (QPTs)~\cite{Amico2008,Eisert2010}.
The entanglement entropy (EE), which quantifies bipartite entanglement (the entanglement between two subsystems), is a typical measure of entanglement~\cite{Bennett1996}.
The signature of EE in QPTs has been studied in various many-body quantum systems, such as, spin system~\cite{Osborne2002,Vidal2003,Latorre2004,Carrasco2016}, Fermi-Hubbard system~\cite{Gu2004,Anfossi2005,Anfossi2006,You2014} and Bose-Hubbard (BH) system~\cite{Buonsante2007,Lauchli2008,Silva-Valencia2011,Pino2012,Alba2013,Irenee2016}.
Recently, the second-order R\'enyi EE in a bosonic superfluid-insulator (SI) transition~\cite{Jaksch1998,Greiner2002} has been measured in an experiment of ultracold bosonic atoms in optical lattices~\cite{Islam2015}.
However, a biggest challenge is to efficiently and rapidly extract the signature of EE for QPTs in an interacting many-body quantum system.

Up to now, several different methods have been devoted to calculate the EE in various BH systems.
For one-dimensional BH systems, the von Neumann entropy (i.e. the first-order R\'enyi EE) has been calculated by exact diagonalization~\cite{Buonsante2007}, density matrix renormalization group (DMRG)~\cite{Lauchli2008}, time-evolving block decimation (TEBD)~\cite{Pino2012} and slave-boson approach~\cite{Irenee2016}.
For two-dimensional BH systems, the R\'enyi EE up to second-order has been calculated by DMRG~\cite{Alba2013}.
For completely connected graphs, the von Neumann entropy has been calculated by exact diagonalization~\cite{Buonsante2007,Giorda2004}.
For a Kagom\'e lattice, the topological EE has been calculated by quantum Monte Carlo (QMC) simulation~\cite{Isakov2011}.
Most of these works concentrate on the von Neumann entropy, the first-order R\'enyi EE, which is hard to be measured in experiments.
Moreover, although the von Neumann entropies for infinite systems may successfully identify QPTs, their calculations request large computational resources and long computational times.
As the second-order R\'enyi EE is experimentally measurable, it is vital to calculate such a measurable EE via an efficient method, which does not cost too much computational resources and times.

It is well-known that the mean-field (MF) approach may successfully predict the emergence of most QPTs in BH systems.
However, the conventional single-site MF approach~\cite{Fisher1989,Sheshadri1993,Oosten2001}, which is based upon the product state ansatz of single-site states, does not include any entanglement.
Therefore, it is impossible to extract EE via the conventional single-site MF approach.
Fortunately, the cluster MF approach, or more generally, the composite boson MF theory~\cite{Huerga2013}, can reserve partial entanglement and have explored several new phases not found by the conventional single-site MF approach~\cite{Buonsante2004,Pisarski2011,McIntosh2012,Luhmann2013,Deng2015}.
Can the cluster MF approach efficiently capture the signature of EE in the emerged QPTs?

In this article, we explore the cluster MF signature of EE for the bosonic SI transitions in a trimerized Kagom\'e lattice (TKL).
In the framework of our cluster MF treatment, each trimer in the TKL is treated as a cluster and the inter-trimer hopping is decoupled by using the MF approximation.
Therefore the ground-state (GS) of the whole system is a product of single-cluster states.
By diagonalizing the MF Hamiltonian, we obtain its GS through a standard self-consistent procedure.
By calculating the order parameter and the filling number, we find that the GS phase diagram include three different types of phases: superfluid, integer insulator (IntI) and fractional insulator (FracI).
By analyzing the second-order R\'enyi EE, we find that the discontinuousness of its first-order derivative corresponds to the critical point of the bosonic SI transitions.
Our results clearly show that the cluster MF treatment can still efficiently capture the signature of EE in the bosonic SI transitions.
As the cluster MF treatment have successfully predicted the appearance of most QPTs in quantum superlattice models, we believe it may also efficiently capture the signature of EE in those QPTs.

The article structure is as follows.
In this section, we introduce the related backgrounds and our motivation.
In Sec.~\ref{Sec2}, we give the Hamiltonian for our physical system.
In Sec.~\ref{Sec3}, we present our cluster MF treatment and obtain the ground-state phase diagram.
In Sec.~\ref{Sec4}, we calculate the second-order R\'enyi EE and its first-order derivative and then show the relation between the first-order derivative and the phase transition.
In Sec.~\ref{Sec5}, we discuss the experimental possibility.
In the last section, we conclude and discuss our results.

\section{model\label{Sec2}}

\begin{figure}[!htp]
  \includegraphics[width=1.0\columnwidth]{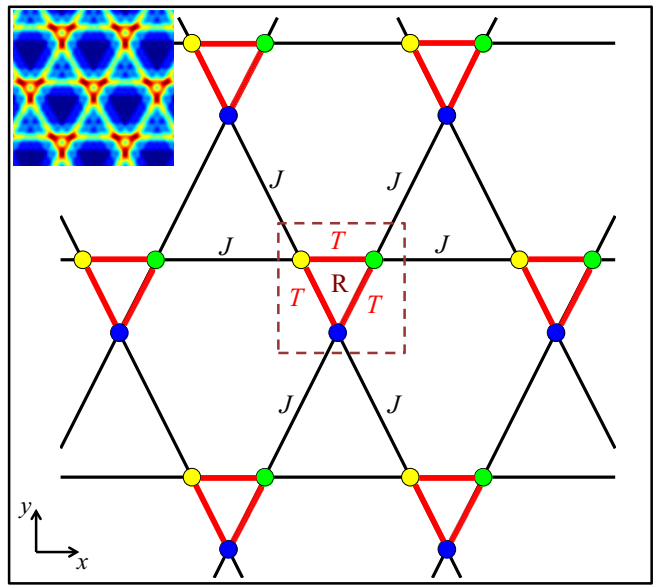}
  \caption{\label{Fig_kagome_lattice}(color online).
  The schematic diagram for the trimerized Kagom\'e lattice.
  The unit cell is labelled by R.
  The lattice sites in the same color are equivalent.
  The red and the black bonds denote the intra-cell hopping $T$ and the inter-cell hopping $J$, respectively.
  The inset shows optical lattice potential, in which the darkest red region denotes the potential minimum.}
\end{figure}

We consider an ensemble of interacting ultracold Bose atoms confined in a two-dimensional optical superlattice,
\begin{eqnarray}
 V(\mathbf r)=V_0\sum_{i=1}^3&\Big[&\cos\left(\mathbf k_i\cdot\mathbf r+\tfrac{3\sigma_i\phi}{2}\right)+2\cos\left(\tfrac{\mathbf k_i\cdot\mathbf r}{3}+\tfrac{\sigma_i\phi}{2}\right) \nonumber\\
  &&+4\cos\left(\tfrac{\mathbf k_i\cdot\mathbf r}{9}+\tfrac{\sigma_i\phi}{6}\right)\Big]^2,
\end{eqnarray}
where $\mathbf k_1=k(\frac{1}{2},\frac{\sqrt{3}}{2})$, $\mathbf k_2=k(\frac{1}{2},-\frac{\sqrt{3}}{2})$, $\mathbf k_3=k(-1,0)$, and $\sigma_1=-\sigma_2=\sigma_3=1$.
The potential depth $V_0$ is proportional to the laser intensity.
The degree of trimerization can be tuned by varying the phase $\phi$~\cite{Santos2004,Damski2005,Lee2006}.
The above potential forms a TKL lattice, see the inset of Fig.~\ref{Fig_kagome_lattice}.

The atom-atom interaction is described by the contact interaction
\begin{equation}
 U_c(\mathbf r-\mathbf r')=\frac{4\pi\hbar^2 a_s}{m}\delta(\mathbf r-\mathbf r')=g\delta(\mathbf r-\mathbf r'),
\end{equation}
where $m$ is the atomic mass and $a_s$ is the $s$-wave scattering length.
In the second quantization theory, the system obeys the Hamiltonian
\begin{eqnarray}
  \hat H &=&\int\mathrm d\mathbf r\hatd \psi(\mathbf r)\left[-\frac{\hbar^2 \nabla^2}{2m}+V(\mathbf r) -\mu\right] \hat \psi(\mathbf r) \nonumber\\
  &+& \frac{1}{2}\int\mathrm d\mathbf r\mathrm d\mathbf r'\hatd \psi(\mathbf r)\hatd \psi (\mathbf r') U_c(\mathbf r-\mathbf r') \hat \psi(\mathbf r') \hat \psi(\mathbf r),
\end{eqnarray}
with the chemical potential $\mu$ and the bosonic field operator $\hat \psi(\mathbf r)$.
If the potential is sufficiently deep, the atoms will only occupy the lowest band and the field operator in terms of the lowest-band Wannier functions, the above Hamiltonian can be simplified as
\begin{eqnarray}\label{Eq.kagome_Ham}
  \hat H_\mathrm{BH}&=&-T\sum_{\langle \mathrm Ri, \mathrm Rj\rangle}{\hatd a_{\mathrm {R}i}\hat a_{\mathrm {R}j}}-J \sum_{\substack{\langle \mathrm{R}i,\mathrm{R'}j\rangle,\mathrm R'\ne \mathrm{R}}}{\hatd a_{\mathrm {R}i} \hat a_{\mathrm {R'}j}} \nonumber\\
  && +\frac{U}{2}\sum_{\mathrm{R}i}{\hat n_{\mathrm {R}i}(\hat n_{\mathrm {R}i}-1)}-\mu\sum_{\mathrm{R}i}{\hat n_{\mathrm {R}i}},
\end{eqnarray}
with $\hatd a_{\mathrm{R}i}$($\hat a_{\mathrm{R}i}$) creating (annihilating) a boson at the $i$-th site of the $\mathrm{R}$-th trimer. In Fig.~\ref{Fig_kagome_lattice}, the lattice sites in blue, yellow and green are labelled as 1, 2 and 3, respectively.
Here, $\hat n_{\mathrm{R}i}=\hatd a_{\mathrm{R}i} \hat a_{\mathrm{R}i}$ is the number operator,
$T$ and $J$ are respectively the intra- and inter-trimer hopping strengthes, $U$ is the on-site interaction and $\mu$ is the chemical potential.
The summations of $\mathrm{R}i$ and $\langle{\mathrm{R}i,\mathrm{R'}j}\rangle$ comprise all lattice sites in the whole system and all nearest neighbor sites in different trimers, respectively.
For convenience, we define $\beta=J/T$, the ratio between inter- and intra-trimer hopping strengths.
Here we concentrate on the trimerized system of repulsive interaction, i.e., $\beta < 1$ and $U>0$.

\section{cluster mean-field phase diagram\label{Sec3}}

In this section, we show how to obtain the ground-state phase diagram with the cluster mean-field approach.
In the first subsection, we present our cluster mean-field treatment for determining the ground states.
In the last subsection, we give the ground-state phase diagram by calculating the order parameters.

\subsection{Cluster mean-field treatment\label{sub1}}

The conventional single-site MF approach decouples the whole lattice into single sites which interact with the surrounding sites through mean fields.
Based upon the equivalence of all single sites, the mean fields of the surrounding sites is thus replaced by the mean field of the site itself.
Thus, the MF version of the original Hamiltonian can be viewed as a sum of single-site Hamiltonians.

The cluster MF approach we use is an extension of the single-site MF approach.
It decouples the whole system as clusters of multiple lattice sites, in which the inter-cluster coupling are decoupled and the intra-clulster coupling are kept unchanged.
In the TKL, we regard each trimer as a cluster and implement the standard MF decoupling for different trimers,
\begin{eqnarray}\label{Eq.decouple}
 \hatd a_{\mathrm Ri} \hat a_{\mathrm R'j} &\rightarrow& \langle \hatd a_{\mathrm Ri} \rangle \hat a_{\mathrm R'j} +\hatd a_{\mathrm Ri} \langle \hat a_{\mathrm R'j} \rangle -\langle \hatd a_{\mathrm Ri} \rangle \langle \hat a_{\mathrm R'j} \rangle.
\end{eqnarray}
By introducing the order parameter $\Psi_{\mathrm{R}i}=\langle \hat a_{\mathrm Ri} \rangle$ and substituting Eq.~\eqref{Eq.decouple} into Eq.~\eqref{Eq.kagome_Ham}, we obtain the MF Hamiltonian
\begin{equation}
  \hat H_\mathrm{MF}=\sum_\mathrm{R} \hat H_\mathrm{MF}^\mathrm{R},
\end{equation}
with the single-cluster MF Hamiltonian
\begin{eqnarray}\label{Eq.cluster_Ham}
 \hat H_\mathrm{MF}^\mathrm{R}\!=&-&T\sum_{\langle i,j\rangle}{\hatd a_{\mathrm{R}i}\hat a_{\mathrm{R}j}} \nonumber\\ &-&J\sum_{\substack{\langle \mathrm{R'}j,\mathrm{R}i\rangle\\ \mathrm R'\ne \mathrm R}}[\hatd a_{\mathrm {R}i}\Psi_{\mathrm {R'}j}+\hat a_{\mathrm{R}i}\Psi^*_{\mathrm{R'}j}-\mathrm{Re}(\Psi_{\mathrm{R}i}^*\Psi_{\mathrm{R'}j})] \nonumber\\ &+&\frac{U}{2}\sum_{i}{\hat n_{\mathrm{R}i}(\hat n_{\mathrm{R}i}-1)}-\mu\sum_{i}{\hat n_{\mathrm{R}i}}.
\end{eqnarray}
The whole GS can be expressed as a product of the single-cluster states,
\begin{equation}\label{Eq.ground_state}
  \ket{\Phi_\mathrm{MF}}=\otimes_\mathrm{R}\ket{\Phi_\mathrm{MF}^\mathrm{R}}.
\end{equation}
Thus, the eigenequation $\hat H_{\mathrm {MF}}\ket {\Phi_{\mathrm {MF}}}=E\ket {\Phi_{\mathrm {MF}}}$ of the whole system can be reduced to the single-cluster eigenequation
\begin{eqnarray}\label{Eq.single_cluster_eigeq}
 \hat H_{\mathrm {MF}}^{\mathrm R}\ket {\Phi_{\mathrm {MF}}^{\mathrm R}}=E_{\mathrm R}\ket {\Phi_{\mathrm {MF}}^{\mathrm R}}
\end{eqnarray}
with $E=\sum_{\mathrm R}E_{\mathrm R}$.

To diagonalize the single-cluster MF Hamiltonian, we introduce the Fock bases $\ket {n_1n_2n_3}_{\mathrm {R}}=\ket {n_1}_{\mathrm R1}\otimes \ket {n_2}_{\mathrm R2}\otimes \ket {n_3}_{\mathrm R3}$ with $n_i$ being the particle number at the $i$-th site with $i=\{1, 2, 3\}$.
In our calculation, the particle number is truncated at $N_{max}$, that is, $0\leq n_i \leq N_{max}$.
Similar to the conventional MF approach, we perform a self-consistent procedure to obtain the single-cluster GS,
\begin{eqnarray}\label{Eq.cluster_ansatz}
 \ket{\Phi_\mathrm{MF}^\mathrm{R}}=\sum_{n_1,n_2,n_3=0}^{\mathrm{N}_{max}}C_{n_1n_2n_3}^{\mathrm{R}}\ket {n_1n_2n_3}_{\mathrm{R}},
\end{eqnarray}
where $C_{n_1n_2n_3}^{\mathrm{R}}$ are the complex amplitudes.

We note that the order parameters at all sites are equivalent for the following reasons.
The Kagom\'e lattice possesses translational symmetry in the sense that it remains unchanged when translate the clusters along $(1,0)$, and $(\cos \frac{\pi}{3}, \sin \frac{\pi}{3})$ direction.
That is to say, all clusters are equivalent.
In addition, after rotating the lattice for $60^\circ$ and $120^\circ$ around the axes, which are through the center of a cluster and perpendicular to the lattice plane, the lattice keeps unchanged.
This means that the three sites in the same cluster are equivalent.
Thus, the three order parameters for the ground state are always equal, $\Psi_{\mathrm Ri}=\Psi_{\mathrm R'j}=\Psi$.

The self-consistent procedure for determining the ground state includes the following key steps:

(\romannumeral1) Initialize $\Psi=0,\Psi'=\Psi$, and substitute $\Psi$ into the single-cluster Hamiltonian $\hat H_{\mathrm {MF}}^\mathrm R$ to obtain the ground-state energy $E_{\mathrm {GS}}^{\mathrm {min}}$.

(\romannumeral2) Let $\Psi=\Psi+\Delta \Psi$, and substitute it into $\hat H_{\mathrm {MF}}^\mathrm R$ to obtain the ground-state energy $E_{\mathrm {GS}}$.

(\romannumeral3) If $E_{\mathrm {GS}}<E_{\mathrm {GS}}^{\mathrm {min}}$, replace $\Psi'$ and $E_{\mathrm {GS}}^{\mathrm {min}}$ with $\Psi$ and $E_{\mathrm {GS}}$ respectively. Otherwise, turn to step (\romannumeral4).

(\romannumeral4) Repeat step (\romannumeral2) and (\romannumeral3) until $\Psi \geqslant \sqrt {N_{\mathrm {max}}}$.

(\romannumeral5) Substitute $\Psi'$ into $\hat H_{\mathrm {MF}}^\mathrm R$ and obtain the GS, $\ket {\mathrm {GS}'}$.

(\romannumeral6) Calculate the order parameter $\Psi''=\bra {\mathrm {GS}'}\hat a_i\ket {\mathrm {GS}'}$.

(\romannumeral7) Compare $\Psi'$ and $\Psi''$.
If $|\Psi'-\Psi''|<\epsilon$ ($\epsilon$ is the given tolerance), then output $\ket {\mathrm {GS}'}$ and $\Psi''$ as the GS and the order parameter respectively.
Otherwise, set $\Psi'=\Psi''$ and return to step (\romannumeral5).

\subsection{Ground-state phase diagram\label{sub2}}

\begin{figure}[!htp]
  \includegraphics[width=1.0\columnwidth]{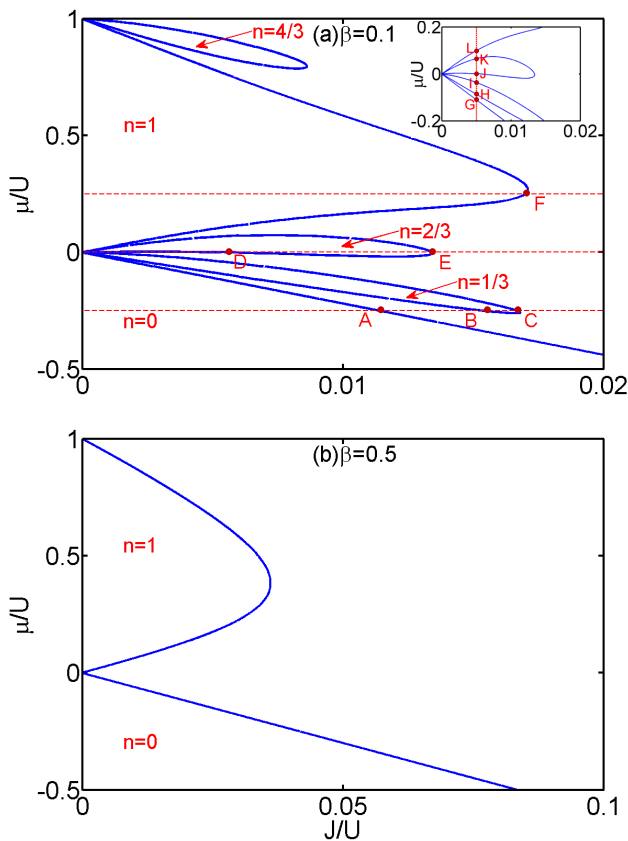}
  \caption{\label{Fig_phase_boundary} The ground-state phase diagram for the bosonic trimerized Kagom\'e lattice with different values of $\beta=J/T$: (a) $\beta=0.1$, (b) $\beta=0.5$.
  In our calculation, we choose the truncation number $N_{max}=8$.
  The insulator lobes with zero order parameters are surrounded by the superfluid phase with nonzero order parameters.
  In (a) (main panel), between neighboring integer insulator lobes, there appear exotic insulator loopholes with fractional filling numbers $n=k/3$ (where $k$ are positive integers).
  The inset in (a) is the magnification of the phase diagram for $\mu/U \in [-0.2,0.2]$.
  }
\end{figure}

In this subsection, we show the phase diagram.
To distinguish different GS phases, we calculate the order parameter $\Psi$ and the filling number $n=\langle \hat{n}_{i}\rangle$.
If $\beta \sim 1$, i.e. $J \sim T$, there are two typical phases in our bosonic TKL: (i) the IntI phase with $\Psi=0$ and integer filling number $n$, and (ii) the superfluid phase with $\Psi \ne 0$.
The superfluid phase appears when the system is dominated by the hopping terms.
Otherwise, when the on-site interaction terms dominate the system, the IntI phase appears.
The filling number $n$ is controlled by the chemical potential $\mu$.
The phase diagram in this case is almost the same as the one for the square BH lattice, see Fig.~\ref{Fig_phase_boundary}(b) for $\beta=0.5$.

If the hopping is dominated by the intra-trimer hopping (i.e. $\beta \ll 1$), there appears a new exotic phase, the FracI phase, which has zero order parameter $\Psi$ and fractional filling number $n$.
In the strong hopping regime ($J/U \rightarrow \infty$), the GS is a superfluid.
In the strong interaction regime ($J/U \rightarrow 0$), dependent on the chemical potential $\mu$ and the ratio $J/U$, the GS is a superfluid, an IntI or a FracI.
As both integer and fractional insulators have zero order parameter, one has to calculate the filling number to distinguish them.
In our bosonic TKL, the filling number of the FracI is multiplies of one third.
Moreover, in the FracI phase, the particles can still move freely in the same cluster although they can not move between different clusters.
In the phase diagram, the FracI phases appear as loopholes between neighboring IntI lobes.
In addition, the loophole domains are obtained in the bosonic TKL by cell strong-coupling perturbation technique\cite{Buonsante2005}, which solves the intra-trimer Hamiltonians exactly and treats the inter-trimer hopping as perturbation.

In Fig.~\ref{Fig_phase_boundary}(a), we show the phase diagram for the system of $\beta=0.1$.
In the main panel of Fig.~\ref{Fig_phase_boundary}(a), the three red-dashed lines are parallel to the $J/U$ axis corresponding to $\mu/U= -0.25$, $0$, and $0.25$, respectively.
Along the line of $\mu/U=-0.25$, the system undergoes the vacuum-superfluid-FracI-superfluid transition with three critical points (A, B, C) at $J/U = (0.01142, 0.01528, 0.01670)$.
Along the line of $\mu/U=0$, the system undergoes the superfluid-FracI-superfluid transition with two critical points (D, E) at $J/U = (0.00548, 0.01340)$.
Along the line of $\mu/U=0.25$, the system undergoes the IntI-superfluid transition with the critical F at $J/U = 0.01702$.
We show in the inset the magnification of the phase diagram depicted in the main panel.
The dashed line is parallel to the $\mu/U$ axis corresponding to $J/U=0.005$.
Along this line, the system undergoes the vacuum-superfluid-FracI-superfluid-FracI-superfluid-IntI transition with the six critical points labeled as (G,H,I,J,K,L) at $\mu/U= (-0.111,-0.087,-0.037, 0.001,0.065,0.099)$.
The phase transitions along such lines, which are driven by the ratio $\mu/U$ or $J/U$ at variable filling, belong to the class of commensurate-incommensurate (CI) transitions.
In addition to the CI transitions, there is another typical class of phase transitions in the BH systems, the O(2) transitions, which are driven by the ratio $J/U$ at fixed integer filling.

\section{entanglement entropy and phase transitions\label{Sec4}}

Now we show how to extract the signature of EE for QPTs in our bosonic TKL.
As mentioned above, the R\'enyi EE is a bipartite entanglement defined through separating the whole system into two subsystems and its second-order form can be measured in experiments.
Denoting the two subsystems as A and B, the $n$-th order R\'enyi EE is defined as
\begin{eqnarray}
 S_n[\mathrm A(\mathrm B)]=\frac {1}{1-n}\log \mathrm{Tr}(\hat \rho _{\mathrm A(\mathrm B)}^n).
\end{eqnarray}
Here, $\hat \rho_{\mathrm A (\mathrm B)}=\mathrm{Tr}_{\mathrm B (\mathrm A)}(\hat \rho_{\mathrm {AB}})$ is the reduced density matrix of subsystem $\mathrm A (\mathrm B)$ and $\hat \rho_{\mathrm {AB}}$ is the density matrix of the whole system.
If the two subsystems are entangled, ignoring information about one subsystem will result in the other subsystem being in a mixed quantum state.
Through implementing the L'H$\hat{o}$pital's rule, it is easy to find that the first-order R\'enyi EE just gives the von Neumann entropy.
Below we concentrate on the second-order ($n=2$) R\'enyi EE, $S_2[\mathrm A(\mathrm B)]=-\log \mathrm{Tr}(\hat \rho_{\mathrm A(\mathrm B)}^2)$.

In our cluster MF treatment, only the intra-cluster correlations are reserved, thus we only need to consider intra-cluster bipartite entanglement.
For a given trimer-cluster R, the subsystem A can be chosen as one of the three sites and the subsystem B are then the rest two sites.
Therefore, the reduced density matrices read as
\begin{equation}
 \hat \rho_{\mathrm A(\mathrm B)}=\mathrm {Tr}_{jk(i)}\hat \rho_\mathrm R,
\end{equation}
where $\{i, j, k\}$ label the three sites in cluster R.
From the single-cluster GS Eq.~\eqref{Eq.cluster_ansatz}, the reduced density matrix $\hat \rho_A$ and $\hat \rho_B$ are given as
\begin{eqnarray}\label{Eq.reduced_density_matrix}
 \hat \rho_{\mathrm A} &=&\sum_{\substack{n_1,n'_1}}\left( \sum_{n_2,n_3}C_{n_1 n_2 n_3}C_{n'_1n_2n_3}^*\right) \ket {n_1} \bra {n'_1}, \nonumber\\
 \hat \rho_{\mathrm B} &=&\sum_{\substack{n_2,n_3 \\ n'_2,n'_3}} \left(\sum_{n_1} C_{n_1 n_2 n_3}C_{n_1n'_2n'_3}^*\right)\ket {n_2 n_3} \bra {n'_2 n'_3}.
\end{eqnarray}
According to the above reduced density matrices, we calculate the 2nd-order R\'enyi EE for different parameters.
Our results show that, the second-order R\'enyi EE itself is continuous everywhere, but the continuousness of its first-order derivative breaks down at the phase boundary.
This means that the QPTs in our bosonic TKL correspond to the jumps in the first-order derivative of the second-order R\'enyi EE.

\begin{figure}[!htp]
  \includegraphics[width=1.0\columnwidth]{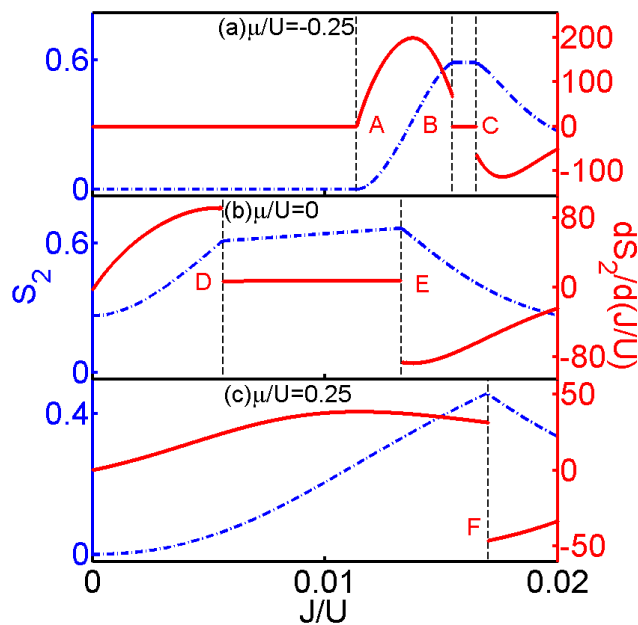}
  \caption{\label{Fig_entropy_der}(color online). The second-order R\'enyi entanglement entropy $S_2$ and its first-order derivative with respect to $J/U$ for different values of $\mu/U$: (a) $-0.25$, (b) $0$, and (c) $0.25$.
  The blue-dot-and-dash and red lines denote $S_2(A)$ and $dS_2/d(J/U)$, respectively.
  The vertical black-dashed lines $\{\mathrm {A, B, C, D, E, F}\}$, where the derivative $dS_2/d(J/U)$ jumps, correspond to the transition points $\{\mathrm {A, B, C, D, E, F}\}$ in the main panel in Fig.~2(a).}
\end{figure}

\begin{figure}[!htp]
  \includegraphics[width=1.0\columnwidth]{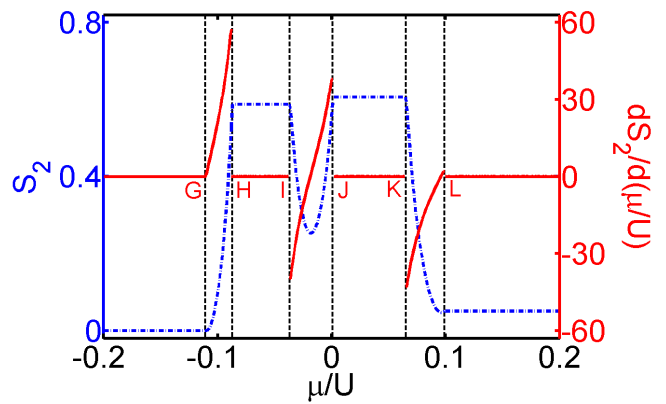}
  \caption{\label{Fig_entropy_mu}(color online). The second-order R\'enyi entanglement entropy $S_2(A)$ and its first-order derivative with respect to $\mu/U$ for $J/U=0.005$.
  The blue-dot-and-dash and red lines denote $S_2$ and $dS_2/d(\mu/U)$, respectively.
  The vertical black-dashed lines $\{\mathrm {G, H, I, J, K, L}\}$, where the derivative $dS_2/d(J/U)$ jumps, correspond to the transition points $\{\mathrm {G, H, I, J, K, L}\}$ in the inset in Fig.~2(a).}
\end{figure}

In Fig.~\ref{Fig_entropy_der}, corresponding to the three red-dashed lines in the main panel of Fig.~\ref{Fig_phase_boundary}(a), we show the second-order R\'enyi EE $S_2$ (blue-dot-and-dash lines) and its first-order derivative with respect to $J/U$ (red lines).
The second-order R\'enyi EE is continuous everywhere but turnpoints appear at the phase boundary.
To characterize the turnpoints, we numerically calculate its first-order derivative with respect to $J/U$ through difference quotient.
Indeed, the first-order derivatives (red lines) show discontinuousness at the phase boundaries (vertical black-dashed lines).
The three vertical black-dashed lines (A, B, C) correspond to the critical points of the vacuum-superfluid-FracI-superfluid transition, see Fig.~\ref{Fig_entropy_der}(a).
The two vertical black-dashed lines (D, E) correspond to the critical points of the superfluid-FracI-superfluid transition, see Fig.~\ref{Fig_entropy_der}(b).
The vertical black-dashed line F corresponds to the critical point of the IntI-superfluid transition, see Fig.~\ref{Fig_entropy_der}(c).
Similarly, the first-order derivative of $S_2$ with respect to $\mu/U$ also shows discontinuousness at the phase boundary, see Fig.~\ref{Fig_entropy_mu}.

The above analysis is based on the class of CI transition.
By employing the slave-boson approach, it has been demonstrated that the von Neumann EE develops cusp singularities at the SI transition in a square BH lattice~\cite{Irenee2016}, which is a CI transition.
Due to that the ``Higgs''-like modes are gapless at the O(2) transitions while they are gapped at the CI transitions, it has been found that the singularities at the O(2) transitions are greatly enhanced~\cite{Irenee2016}.
So, we believe that our findings may apply to the O(2) transitions as well.

\section{experimental possibility\label{Sec5}}

Based upon the present techniques of manipulating and detecting ultracold atoms in optical lattices, it is possible to realize our bosonic TKL and measure its second-order R\'enyi EE.
By imposing three standing-wave lasers, the Kagom\'e optical lattice has been realized and our bosonic TKL can be realized by loading ultracold Bose atoms in the the Kagom\'e optical lattice~\cite{Jo2012}.
Based upon the quantum interference between many-body twins~\cite{Daley2012}, the experimental measurement of the second-order R\'enyi EE has been demonstrated by using simple BH systems~\cite{Islam2015}.
Similarly, the second-order R\'enyi EE in our bosonic TKL can be measured through three steps: state initialization, quantum interference and parity readout.

\begin{figure}[!htp]
 \includegraphics[width=1.0\columnwidth]{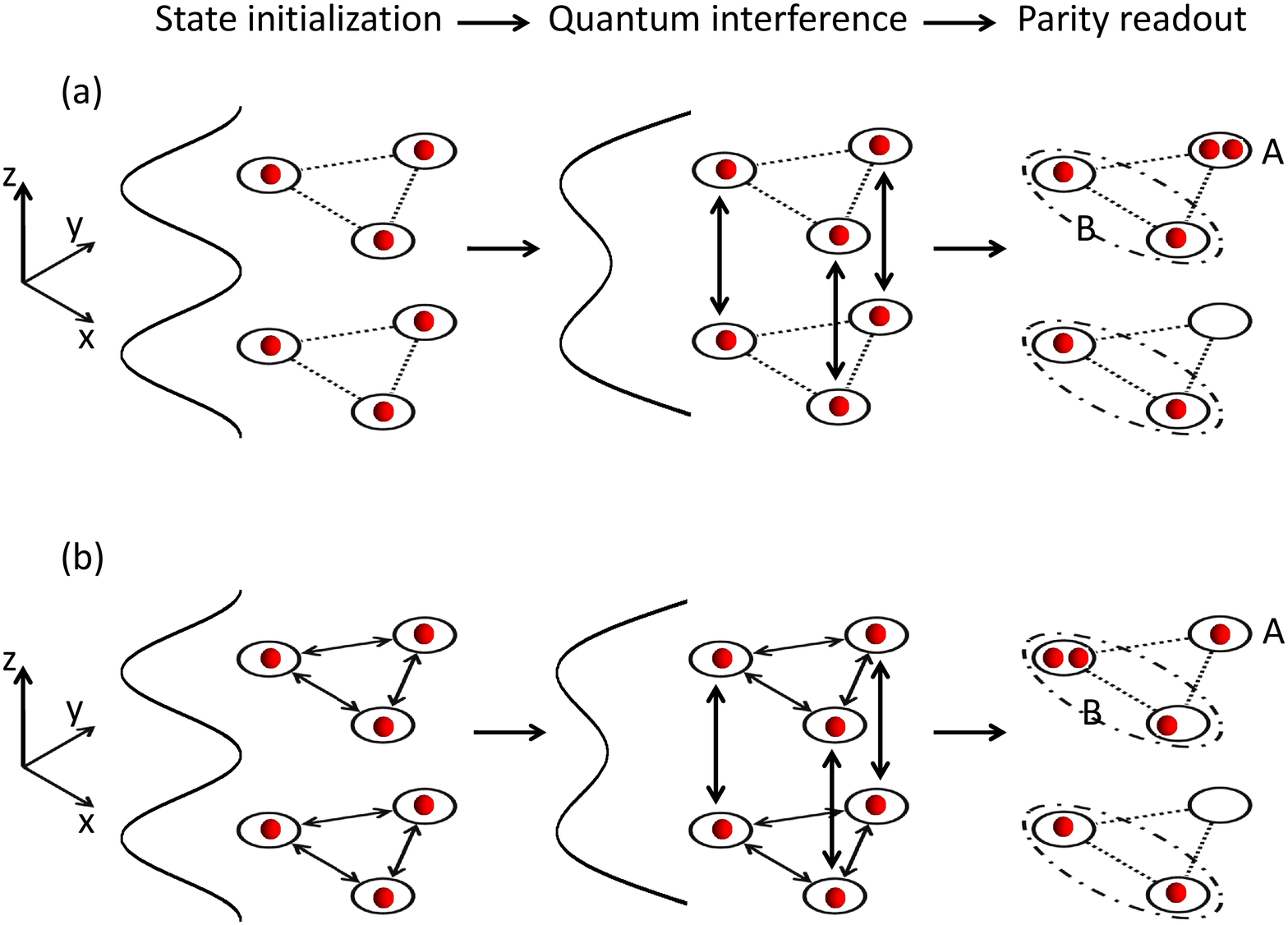}
  \caption{\label{Fig_experimental_setup}(color online). Schematic diagram for measuring the second-order R\'enyi entanglement entropy in our bosonic TKL via quantum interference of many-body twins. The cluster (a trimer) is divided into two subsystems A and B.
  The measurement procedure includes three steps: state initialization, quantum interference and parity readout.
  (a) For a product state of the two subsystems, which has zero second-order R\'enyi EE, the final particle numbers of the subsystems and the whole system are even.
  (b) For an entangled state of the two subsystems, which has nonzero second-order R\'enyi EE, the final particle numbers in the subsystems are even or odd while they are even in the whole system.}
\end{figure}

The schematic diagram for the measurement is shown in Fig.~\ref{Fig_experimental_setup}.
Firstly, prepare two identical clusters in two decoupled parallel TKLs and initialize them into the same state.
This can be achieved by superimposing a box potential onto a three-dimensional TKL to decouple a bilayer TKL~\cite{Jo2012} and then applying a proper double-well potential to decouple two identical clusters in different layers.
Secondly, perform a discrete Fourier transformation on the two copies.
This can be realized by the beam-splitter operation through lowing the potential barrier between the two layers, in which the bosons in different layers interfere through inter-layer quantum tunneling.
Finally, read out the number of particles in one copy and its subsystems through site-resolved readout.
From the measured total particle number in subsystem A, $n_{\mathrm A}=\sum_{i\in \mathrm A}n_i$, one can obtain the purity $\mathrm {Tr}(\rho_{\mathrm A}^2)=\langle(-1)^{n_{\mathrm A}}\rangle$, which is just the average measured parity $P_{\mathrm A}=(-1)^{n_{\mathrm A}}$.
Through the parity measurement of subsystem A, the second-order R\'enyi EE is given as $S_2(\mathrm A)=-\log \mathrm {Tr}(\rho_{\mathrm A}^2)=-\log {\langle (-1)^{n_{\mathrm A}}\rangle}$.
Thus, $S_2$ is zero if A and B are not entangled and nonzero if A and B are entangled.

\section{conclusion and discussion\label{Sec6}}

In summary, based upon the cluster MF treatment, we have successfully extract the signature of second-order R\'enyi EE for the QPTs of interacting bosons in TKL.
In addition to the superfluid and integer insulator phases, we find that fractional insulator phase may appear in strongly trimerized systems.
Through analyzing the second-order R\'enyi EE and its first-order derivative, we explore that, while the second-order R\'enyi EE itself is continuous everywhere, its first-order derivative becomes discontinuous at the quantum critical points.
This means that the intra-cluster entanglement reserved in the cluster MF treatment is still sufficient for capturing the signature of EE in QPTs.

Our method for extracting the residual intra-cluster entanglement provides a highly efficient and rapid way to capture the signature of EE for the QPTs in quantum superlattice models.
In comparison with the TEBD, DMRG and QMC methods~\cite{Lauchli2008,Pino2012,Alba2013,Isakov2011}, which cost huge computational resources and long computational times, the calculation of EE via the cluster MF treatment costs much less computational resources and short computational time.
In comparison with the exact diagonalization~\cite{Buonsante2007}, which can only give phase crossovers, the calculation of EE via the cluster MF treatment may explore true phase transitions.
As the cluster MF approach is a powerful tool for exploring most QPTs in various many-body quantum models (from Fermi-Hubbard model, Bose-Hubbard model to Heisenberg spin model) with different superlattice configurations (from double-well lattices\cite{Sebby-Strabley2006,Chen2011,Atala2014,Tokuno2014,Keles2015,Deng2015}, Kagom\'e lattices\cite{Mielke1992,Santos2004,Damski2005,Jo2012,Pudleiner2015} to honeycomb lattices\cite{Grynberg1993,Snoek2007,Lee2009,Lembessis2015}), beyond extracting the signature of EE for bosonic SI transitions, we believe that our analysis can be easily extended to extract the signature of EE for other QPTs in quantum superlattice systems.

We also note that the cluster dynamical mean-field approach (CDMFA) has been employed to study the entanglement spectrum of the half-filling Hubbard system~\cite{Udagawa2015}.
Although both the cluster mean-field approach (CMFA) and the CDMFA can be used to decouple many-body lattice model into many-body cluster model, they are different. 
The CDMFA based upon the dynamical mean-field theory freezes spatial fluctuations but takes full account of local temporal fluctuations. 
The CMFA we use here takes into account first-order spatial fluctuations but ignores temporal fluctuations.

\acknowledgments{This work was supported by the National Basic Research Program of China (Grant No. 2012CB821305) and the National Natural Science Foundation of China (Grants No. 11374375 and No. 11574405).}


\end{document}